\documentclass[9pt,pdftex,a4paper]{article}%
\pdfoutput=1
 \usepackage{enumitem}
\setlist{nolistsep}
\usepackage{microtype}

\usepackage{amssymb}
\usepackage{multicol}
\usepackage{amsmath}
\usepackage{framed}
\usepackage{float}
\floatstyle{boxed} 
\restylefloat{figure}
\usepackage{tikz}
\usepackage{tabularx}
\usepackage{cleveref}
\crefname{section}{\S}{\S\S}
\usepackage{hyperref}
\usepackage{macros}
\usepackage[ruled]{algorithm}
\usepackage{algorithmicx}
\usepackage{multicol}
\usepackage{caption}

\usepackage{algpseudocode}
\usepackage{ioa_code}
\usepackage{graphicx}
\usepackage[textsize=tiny]{todonotes}
\usepackage{listings}
\usepackage{authblk}
\lstdefinestyle{customc}{
    belowcaptionskip=1\baselineskip,
    breaklines=true,
    frame=L,	
    xleftmargin=\parindent,
    language=C,
    showstringspaces=false,
    escapeinside={//}{\^^M},
    basicstyle=\LSTfont, 
    keywordstyle=\bfseries\color{green!40!black},
    commentstyle=\itshape\color{gray!60!black},
    identifierstyle=\color{blue!50!black},
    stringstyle=\color{orange},
    numbers=left,                    
    numbersep=5pt,                   
    numberstyle=\tiny\color{black},  
    otherkeywords={then,word,process_local,type,xbegin,xabort,xend}
}

\def\S{\ensuremath{\mathcal{S}}}

\newcommand{\RNum}[1]{\uppercase\expandafter{\romannumeral #1\relax}}

\newcommand{\remove}[1]{}

\newtheorem{theorem}{Theorem}

\newcounter{linenumber}

\title{Generalized Paxos Made Byzantine\\ (and Less Complex)}

\usepackage{url}

\author{
Miguel Pires$^1$~~~Srivatsan Ravi$^2$~~~Rodrigo Rodrigues$^1$ \\
$^1$\normalsize INESC-ID and Instituto Superior T\'{e}cnico (U.\ Lisboa), Lisbon, Portugal \\
$^2$\normalsize University of Southern California, Los Angeles, USA
}

\begin{document}

\maketitle

\begin{abstract}
One of the most recent members of the \emph{Paxos} family of protocols is \emph{Generalized} Paxos. 
This variant of Paxos has the characteristic that it departs from the original specification of consensus, allowing for a weaker safety condition where different processes can have a different views on a sequence being agreed upon. 
However, much like the original Paxos counterpart, Generalized Paxos does not have a simple implementation.
Furthermore, with the recent practical adoption of Byzantine fault tolerant protocols, it is timely and important to understand how Generalized Paxos can be implemented in the Byzantine model.
In this paper, we make two main contributions. First, we provide a description of Generalized Paxos that is easier to understand, based on a simpler specification and the pseudocode for a solution that can be readily implemented. Second, we extend the protocol to the Byzantine fault model.
\end{abstract}
%
%


%
\section{Introduction}
\label{sec:intro}
%
The evolution of the Paxos~\cite{Lam98} protocol is an unique
chapter in the history of Computer Science. It was first described in
1989 through a technical report~\cite{paxos:tr}, and was only
published a decade later~\cite{Lam98}. Another long wait took
place until the protocol started to be studied in depth and used by
researchers in various fields, namely the distributed
algorithms~\cite{DPLL97} and the distributed systems~\cite{petal}
research communities. And finally, another decade later, the protocol
made its way to the core of the implementation of the services that
are used by millions of people over the Internet, in particular since
Paxos-based state machine replication is the key component of Google's
Chubby lock service~\cite{chubby}, or the open source ZooKeeper
project~\cite{zookeeper}, used by Yahoo!\ among others. Arguably, the
complexity of the presentation may have stood in the way of a faster
adoption of the protocol, and several attempts have been made at
writing more concise explanations of
it~\cite{L01,Renesse2011}.

More recently, several variants of Paxos have been proposed and
studied. Two important lines of research can be highlighted in this
regard. First, a series of papers hardened the protocol against
malicious adversaries by solving consensus in a Byzantine fault
model~\cite{Martin2006,Lamport2011}. The importance of this line of
research is now being confirmed as these protocols are now in widespread
use in the context of cryptocurrencies and distributed ledger
schemes such as blockchain~\cite{bitcoin}.
Second, many proposals target improving
the Paxos protocol by eliminating communication costs~\cite{L06},
including an important evolution of the protocol called Generalized
Paxos~\cite{Lamport2005}, which has the noteworthy aspect of
having lower communication costs by leveraging a more general specification than traditional consensus that can lead to a weaker requirement in terms of ordering of commands across replicas. In particular, instead of forcing all
processes to agree on the same value, it allows processes to pick an
increasing sequence of commands that differs from process to process
in that commutative commands may appear in a different order.
The practical importance of such weaker specifications is underlined
by a significant research activity on the corresponding weaker consistency
models for replicated systems~\cite{LLS90,dynamo}.


In this paper, we draw a parallel between the evolution of the Paxos
protocol and the current status of Generalized Paxos. In particular,
we argue that, much in the same way that the clarification of the Paxos
protocol contributed to its practical adoption, it is also important
to simplify the description of Generalized Paxos. Furthermore, we believe
that evolving this protocol to the Byzantine model is an important
task, since it will contribute to the understanding
and also open the possibility of adopting Generalized Paxos in
scenarios such as a Blockchain deployment.

As such, the paper makes several contributions, which are listed next.
\begin{itemize}
\item
We present a simplified version of the specification of Generalized
Consensus, which is focused on the most commonly used case of the
solutions to this problem, which is to agree on a sequence of
commands;

\item
we extend the Generalized Paxos protocol to the Byzantine model; 

\item
we present a description of the Byzantine Generalized Paxos protocol
that is more accessible than the original description, namely including the
respective pseudocode, in order to to make it 
easier to implement;

\item
we prove the correctness of the Byzantine Generalize Paxos protocol;

\item
and we discuss several extensions to the protocol in the context of relaxed consistency models and fault tolerance.
\end{itemize}
The remainder of the paper is organized as follows:
Section~\ref{sec:related} gives an overview of Paxos and its family of related protocols.
Section~\ref{sec:model} introduces the model and simplified specification of Generalized Paxos.
Section~\ref{sec:protocol} presents the Generalized Paxos protocol that is resilient against Byzantine failures. Section~\ref{bft_proof} presents a correctness proof for Byzantine Generalized Paxos.
Section~\ref{sec:disc} concludes the paper with a discussion of several optimizations and practical considerations.
%

%

%

\section{Background and related work}
\label{sec:related} 
\subsection{Paxos and its variants} \label{Paxos} 

The Paxos protocol family solves consensus by finding an equilibrium in face of the well-known FLP impossibility result~\cite{FLP85}. It does this by always guaranteeing safety despite asynchrony, but
foregoing progress during the temporary periods of asynchrony, or if more than $f$ faults occur for a system of $N > 2f$ replicas~\cite{L01}. The classic form of Paxos uses a set of proposers, acceptors and learners, runs in a sequence of ballots, and employs two phases (numbered 1 and 2), with a similar message pattern: proposer to acceptors (phase 1a or 2a), acceptors to proposer (phase 1b or 2b), and, in phase 2b, also acceptors to learners. To ensure progress during synchronous periods, proposals are serialized by a distinguished proposer, which is called the leader.\par
Paxos is most commonly deployed as Multi (Decree)-Paxos, which provides an optimization of the basic message pattern by omitting the first phase of messages from all but the first ballot for each leader~\cite{Renesse2011}. This means that a leader only needs to send a \textit{phase 1a} message once and subsequent proposals may be sent directly in \textit{phase 2a} messages. This reduces the message pattern in the common case from five message delays to just three (from proposing to learning). \par
Fast Paxos observes that it is possible to improve on the previous latency (in the common case) by allowing proposers to propose values directly to acceptors~\cite{L06}. To this end, the protocol distinguishes between fast and classic ballots, where fast ballots bypass the leader by sending proposals directly to acceptors and classic ballots work as in the original Paxos protocol. The reduced latency of fast ballots comes at the added cost of using a quorum size of $N-e$ instead of a classic majority quorum, where $e$ is the number of faults that can be tolerated while using fast ballots. In addition, instead of the usual requirement that $N> 2f$, to ensure that fast and classic quorums intersect, a new requirement must be met: $N > 2e+f$. This means that if we wish to tolerate the same number of faults for classic and fast ballots (i.e., $e=f$), then the minimum number of replicas is $3f+1$ instead of the usual $2f+1$. Since fast ballots only take two message steps (\textit{phase 2a} messages between a proposer and the acceptors, and \textit{phase 2b} messages between acceptors and learners), there is the possibility of two proposers concurrently proposing values and generating a conflict, which must be resolved by falling back to a recovery protocol.\looseness=-1 \par
Generalized Paxos improves the performance of Fast Paxos by addressing the issue of collisions. In particular, it allows acceptors to accept different sequences of commands as long as non-commutative operations are totally ordered \cite{Lamport2005}. In the original description, non-commutativity between operations is generically represented as an interference relation. In this context, Generalized Paxos abstracts the traditional consensus problem of agreeing on a single value to the problem of agreeing on an increasing set of values. \textit{C-structs} provide this increasing sequence abstraction and allow the definition of different consensus problems. If we define the sequence of learned commands of a learner $l_i$ as a \textit{c-struct} $learned_{l_i}$, then the consistency requirement for generalized consensus can be defined as: $learned_{l_1}$ and $learned_{l_2}$ must have a \textit{common upper bound}, for all learners $l_1$ and $l_2$. This means that, for any $learned_{l_1}$ and $learned_{l_2}$, there must exist some \textit{c-struct} of which they are both prefixes. This prohibits interfering commands from being concurrently accepted because no subsequent \textit{c-struct} would extend them both. \par
More recently, other Paxos variants have been proposed to address specific issues. For example, Mencius~\cite{Mao2008} avoids the latency penalty in wide-area deployments of having a single leader, through which every proposal must go through. In Mencius, the leader of each round rotates between every process: the leader of round $i$ is process $p_k$, such that $k = n\ mod\ i$.  
Another variant is Egalitarian Paxos (EPaxos), which achieves a better throughput than Paxos by removing the bottleneck caused by having a leader \cite{Moraru2013}. To avoid choosing a leader, the proposal of commands for a command slot is done in a decentralized manner, taking advantage of the commutativity observations made by Generalized Paxos \cite{Lamport2005}. Conflicts between commands are handled by having replicas reply with a command dependency, which then leads to falling back to using another protocol phase with $f+\lfloor\frac{f+1}{2}\rfloor$ replicas.

\subsection{Byzantine fault tolerant replication} \label{Non-Crash}
Consensus in the Byzantine model was originally defined by Lamport et al.~\cite{LSP82}. Almost two decades later, a surge of research in the area started with the PBFT protocol, which solves consensus for state machine replication with $3f+1$ replicas while tolerating up to $f$ Byzantine faults \cite{CL99}. In PBFT, the system moves through configurations called \textit{views}, in which one replica is the primary and the remaining replicas are the backups. The protocol proceeds in a sequence of steps, where messages are sent from the client to the primary, from the primary to the backups, followed by two all-to-all steps between the replicas, with the last step proceeding in parallel with sending a reply to the clients. \par
Zeno is a Byzantine fault tolerance state machine replication protocol that trades availability for consistency~\cite{Singh2009}. In particular, it offers eventual consistency by allowing state machine commands to execute in a \textit{weak quorum} of  $f+1$ replicas. This ensures that at least one correct replica will execute the request and commit it to the linear history, but does not guarantee the intersection property that is required for linearizability. \par
The closest related work is Fast Byzantine Paxos (FaB), which solves consensus in the Byzantine setting within two message communication steps in the common case, while requiring $5f+1$ acceptors to ensure safety and liveness \cite{Martin2006}. A variant that is proposed in the same paper is the Parameterized FaB Paxos protocol, which generalizes FaB by decoupling replication for fault tolerance from replication for performance. As such, the Parameterized FaB Paxos requires $3f+2t+1$ replicas to solve consensus, preserving safety while tolerating up to $f$ faults and completing in two steps despite up to $t$ faults. Therefore, FaB Paxos is a special case of Parameterized FaB Paxos where $t=f$. It has also been shown that $N>5f$ is a lower bound on the number of acceptors required to guarantee 2-step execution in the Byzantine model. In this sense, the FaB protocol is tight since it requires $5f+1$ acceptors to provide the same guarantees.\par
In comparison to the FaB Paxos protocol, our BGP protocol requires a lower number of acceptors than what is stipulated by FaB Paxos' lower bound~\cite{Martin2006}. However, this does not constitute a violation of the result since BGP does not guarantee a two step execution in the Byzantine scenario. Instead, BGP only provides a two communication step latency when proposed sequences are universally commutative with any other sequence. In the common case, BGP requires three messages steps for a sequence to be learned. In other words, Byzantine Generalized Paxos introduces an additional broadcast phase to decrease the requirements regarding the minimum number of acceptor processes. This may be a sensible trade-off in systems that target datacenter environments where communication between machines is fast and a high percentage of costs is directly related to equipment. The fast communication links would mitigate the latency cost of having an additional phase between the acceptors and the high cost of equipment and power consumption makes the reduced number of acceptor processes attractive.\looseness=-1
\raggedbottom
\section{Model}
\label{sec:model}
We consider an \emph{asynchronous} system in which
a set of $n \in \mathbb{N}$ processes communicate by 
\emph{sending} and \emph{receiving} messages.
Each process executes an algorithm assigned to it, but may fail in two different ways. First, it may stop executing it by \emph{crashing}.
Second, it may stop following the algorithm assigned to it, in which case it is considered \emph{Byzantine}. We say that a non-Byzantine process is \emph{correct}.
This paper considers the \emph{authenticated} Byzantine model: every process can produce cryptographic digital signatures~\cite{quorum}. 
Furthermore, for clarity of exposition, we assume authenticated perfect links~\cite{cgr:book}, 
where a message that is sent by a non-faulty sender is eventually received and messages cannot be forged 
(such links can be implemented trivially using retransmission, elimination of duplicates, and point-to-point message authentication codes~\cite{cgr:book}.)
A process may be a \emph{learner}, \emph{proposer} or \emph{acceptor}.
Informally, proposers provide input values that must be agreed upon by learners, the acceptors help the learners \emph{agree} on a value, and learners learn commands by appending them to a local sequence of commands to be executed, $learned_l$ .
Our protocols require a minimum number of acceptor processes ($N$), which is a function of the maximum number of tolerated Byzantine faults ($f$), namely $N \ge 3f+1$. We assume that acceptor processes have identifiers in the set $\{0,...,N-1\}$. In contrast, the number of proposer and learner processes can be set arbitrarily.\looseness=-1\par
\noindent\textbf{Problem Statement.}
In our simplified specification of Generalized Paxos, each learner $l$ maintains a monotonically increasing sequence of commands $learned_l$. 
We define two learned sequences of commands to be equivalent ($\thicksim$) 
if one can be transformed into the other by permuting the elements in a way such that the order of non-commutative pairs is preserved. A sequence $x$ is defined to be an \textit{eq-prefix} of another sequence $y$ ($x \sqsubseteq y$), if the subsequence of $y$ that contains all the elements in $x$ is equivalent ($\thicksim$) to $x$. 
We present the requirements for this consensus problem, stated in terms of learned sequences of commands for a correct learner $l$, $learned_l$. 
To simplify the original specification, instead of using c-structs (as explained in Section~\ref{sec:related}), we specialize to agreeing on equivalent sequences of commands:\par
\begin{enumerate}
\item \textbf{Nontriviality.} If all proposers are correct, $learned_l$ can only contain proposed commands.
\item \textbf{Stability.} If $learned_l = s$ then, at all later times, $s \sqsubseteq learned_l$, for any sequence $s$ and correct learner $l$.
\item \textbf{Consistency.} At any time and for any two correct learners $l_i$ and $l_j$, $learned_{l_i}$ and $learned_{l_j}$ can subsequently be extended to equivalent sequences.
\item \textbf{Liveness.} For any proposal $s$ from a correct proposer, and correct learner $l$, eventually $learned_l$ contains $s$.
\end{enumerate}

\section{Protocol}
\label{sec:protocol}

This section presents our Byzantine fault tolerant Generalized Paxos
Protocol (or BGP, for short). 

\begin{algorithm}
	\caption{Byzantine Generalized Paxos - Proposer p}
	\label{BFT-Prop}
	\textbf{Local variables:} $ballot\_type = \bot$
	\begin{algorithmic}[1]	
		
		\State \textbf{upon} \textit{receive($BALLOT, type$)} \textbf{do} 
		\State \hspace{\algorithmicindent} $ballot\_type = type$;
		\State
		
		\State \textbf{upon} \textit{command\_request($c$)} \textbf{do}   \hspace{\algorithmicindent}\hspace{\algorithmicindent}\hspace{\algorithmicindent}\hspace{\algorithmicindent}
		\State \hspace{\algorithmicindent} \textbf{if} $ballot\_type == fast\_ballot$ \textbf{then}
		\State \hspace{\algorithmicindent}\hspace{\algorithmicindent} $\Call{send}{P2A\_FAST, c}$ to acceptors;
		\State \hspace{\algorithmicindent} \textbf{else} 
		\State \hspace{\algorithmicindent}\hspace{\algorithmicindent} $\Call{send}{PROPOSE, c}$ to leader;		
	\end{algorithmic}
\end{algorithm}

\subsection{Overview}
We modularize our protocol explanation according to the following main components, which are also present in other protocols of the Paxos family:

\begin{itemize}
	
	\item
	{\bf View Change} -- The goal of this subprotocol is to ensure that, at any given moment, one of the proposers is chosen as a distinguished leader, who runs a specific version of the agreement subprotocol. To achieve this, the view change subprotocol continuously replaces leaders, until one is found that can ensure progress (i.e., commands are eventually appended to the current sequence).
	
	\item
	{\bf Agreement} -- Given a fixed leader, this subprotocol extends the current sequence with a new command or set of commands. Analogously to Fast Paxos~\cite{L06} and Generalized Paxos~\cite{Lamport2005}, choosing this extension can be done through two variants of the protocol: using either \textit{classic} ballots or \textit{fast} ballots, with the characteristic that fast ballots complete in fewer communication steps, but may have to fall back to using a classic ballot when there is contention among concurrent requests.
	
\end{itemize}

\subsection{View Change} 

The goal of the view change subprotocol is to elect a distinguished proposer process, called the leader, that carries through the agreement protocol (i.e., enables proposed commands to eventually be learned by all the learners). The overall design of this subprotocol is similar to the corresponding part of existing BFT state machine replication protocols~\cite{CL99}.\par

In this subprotocol, the system moves through sequentially numbered views, and the leader for each view is chosen in a rotating fashion using the simple equation $\textit{leader(view)}=\textit{view mod N}$. The protocol works continuously by having acceptor processes monitor whether progress is being made on adding commands to the current sequence, and, if not, by multicasting a signed {\sc suspicion} message for the current view to all acceptors suspecting the current leader. Then, if enough suspicions are collected, processes can move to the subsequent view. However, the required number of suspicions must be chosen in a way that prevents Byzantine processes from triggering view changes spuriously. To this end, acceptor processes will multicast a view change message indicating their commitment to starting a new view only after hearing that $f+1$ processes suspect the leader to be faulty. This message contains the new view number, the $f+1$ signed suspicions, and is signed by the acceptor that sends it. This way, if a process receives a view-change message without previously receiving $f+1$ suspicions, it can also multicast a view-change message, after verifying that the suspicions are correctly signed by $f+1$ distinct processes.
This guarantees that if one correct process receives the $f+1$ suspicions and multicasts the view-change message, then all correct processes, upon receiving this message, will be able to validate the proof of $f+1$ suspicions and also multicast the view-change message.\par
\begin{algorithm} 
	\caption{Byzantine Generalized Paxos - Leader l}
	\label{BFT-Lead}
	\textbf{Local variables:} $ballot_l = 0,proposals = \bot, accepted = \bot, notAccepted = \bot, view = 0$
	\begin{algorithmic}[1]
		\State \textbf{upon} \textit{receive($LEADER,view_a,proofs$)} from acceptor \textit{a} \textbf{do}
		\State \hspace{\algorithmicindent} $valid\_proofs = 0$;
		\State \hspace{\algorithmicindent} \textbf{for} $p$ \textbf{in} $acceptors$ \textbf{do} 
		\State \hspace{\algorithmicindent}\hspace{\algorithmicindent} $view\_proof = proofs[p]$;
		
		\State \hspace{\algorithmicindent}\hspace{\algorithmicindent} \textbf{if} $view\_proof_{pub_p} == \langle view\_change, view_a \rangle$ \textbf{then}
		\State \hspace{\algorithmicindent}\hspace{\algorithmicindent}\hspace{\algorithmicindent}  $valid\_proofs \mathrel{+{=}} 1$;
		\State \hspace{\algorithmicindent} \textbf{if} $valid\_proofs > f$ \textbf{then}
		\State \hspace{\algorithmicindent}\hspace{\algorithmicindent} $view = view_a$;
		
		\State
		\State \textbf{upon} \textit{trigger\_next\_ballot($type$)} \textbf{do}
		\State \hspace{\algorithmicindent} $ballot_l \mathrel{+{=}} 1$;
		\State \hspace{\algorithmicindent} $\Call{send}{BALLOT,type}$ to proposers;
		\State \hspace{\algorithmicindent} \textbf{if} $type == fast$ \textbf{then}
		\State \hspace{\algorithmicindent}\hspace{\algorithmicindent} $\Call{send}{FAST,ballot_l,view}$ to acceptors;
		\State \hspace{\algorithmicindent} \textbf{else}
		\State \hspace{\algorithmicindent}\hspace{\algorithmicindent} $\Call{send}{P1A, ballot_l, view}$ to acceptors;
		
		\State
		\State \textbf{upon} \textit{receive($PROPOSE, prop$)} from proposer \textbf{do} 
		\State \hspace{\algorithmicindent} \textbf{if} $\Call{isUniversallyCommutative}{prop}$ \textbf{then}
		\State \hspace{\algorithmicindent}\hspace{\algorithmicindent} $\Call{send}{P2A\_CLASSIC, ballot_l,view, prop}$;
		\State \hspace{\algorithmicindent} \textbf{else}
		\State \hspace{\algorithmicindent}\hspace{\algorithmicindent} $proposals = proposals \bullet prop$;
		
		\State
		\State \textbf{upon} \textit{receive($P1B, ballot, bal_a, proven,val_a, proofs$)} from acceptor $a$ \textbf{do}
		\State \hspace{\algorithmicindent} \textbf{if} $ballot \neq ballot_l$ \textbf{then}
		\State \hspace{\algorithmicindent}\hspace{\algorithmicindent} \textbf{return};
		\State
		\State \hspace{\algorithmicindent} $valid\_proofs = 0$; 
		\State \hspace{\algorithmicindent} \textbf{for} $i$ \textbf{in} $acceptors$ \textbf{do}
		\State \hspace{\algorithmicindent}\hspace{\algorithmicindent} $proof = proofs[proven][i]$;
		\State \hspace{\algorithmicindent}\hspace{\algorithmicindent} \textbf{if} $proof_{pub_i} == \langle bal_a, proven \rangle$ \textbf{then}
		\State \hspace{\algorithmicindent}\hspace{\algorithmicindent}\hspace{\algorithmicindent} 
		$valid\_proofs \mathrel{+{=}} 1$;
		\State
		\State \hspace{\algorithmicindent} \textbf{if} $valid\_proofs > N-f$ \textbf{then}
		\State \hspace{\algorithmicindent}\hspace{\algorithmicindent}\hspace{\algorithmicindent} $accepted[ballot_l][a] = proven$;
		\State \hspace{\algorithmicindent}\hspace{\algorithmicindent}\hspace{\algorithmicindent}		$notAccepted[ballot_l] = notAccepted[ballot_l] \bullet (val_a \setminus proven)$;		
		
		\State 
		\State \hspace{\algorithmicindent}\hspace{\algorithmicindent} \textbf{if} $\#(accepted[ballot_l]) \geq N-f$ \textbf{then} 
		\State \hspace{\algorithmicindent}\hspace{\algorithmicindent}\hspace{\algorithmicindent} $\Call{phase\_2a}{ }$;
		
		\State
		\Function{phase\_2a}{$ $}
		\State $maxTried = \Call{largest\_seq}{accepted[ballot_l]}$;
		\State $previousProposals = \Call{remove\_duplicates}{notAccepted[ballot_l]}$;
		\State $maxTried = maxTried \bullet previousProposals \bullet proposals$;
		\State $\Call{send}{P2A\_CLASSIC,ballot_l,view, maxTried}$ to acceptors;
		\State $proposals = \bot$;
		\EndFunction
		
	\end{algorithmic}
\end{algorithm}

Finally, an acceptor process must wait for $N-f$ view-change messages to start participating in the new view (i.e., update its view number and the corresponding leader process). At this point, the acceptor also assembles the $N-f$ view-change messages proving that others are committing to the new view, and sends them to the new leader. This allows the new leader to start its leadership role in the new view once it validates the $N-f$ signatures contained in a single message.

\subsection{Agreement Protocol} 

The consensus protocol allows learner processes to agree on equivalent sequences of commands (according to the definition of equivalence present in Section~\ref{sec:model}). An important conceptual distinction between Fast Paxos~\cite{L06} and our protocol is that ballots correspond to an extension to the sequence of learned commands of a single ongoing consensus instance, instead of being a separate instance of consensus,. Proposers can try to extend the current sequence by either single commands or sequences of commands. We use the term \textit{proposal} to denote either the command or sequence of commands that was proposed.\par
Ballots can either be \textit{classic} or \textit{fast}. In classic ballots, a leader proposes a single proposal to be appended to the commands learned by the learners. The protocol is then similar to the one used by classic Paxos~\cite{Lam98}, with a first phase where each acceptor conveys to the leader the sequences that the acceptor has already voted for (so that the leader can resend commands that may not have gathered enough votes), followed by a second phase where the leader instructs and gathers support for appending the new proposal to the current sequence of learned commands. Fast ballots, in turn, allow any proposer to contact all acceptors directly in order to extend the current sequence (in case there are no conflicts between concurrent proposals). However, both types of ballots contain an additional round, called the verification phase, in which acceptors broadcast proofs among each other indicating their committal to a sequence. This additional round comes after the acceptors receive a proposal and before they send their votes to the learners.\par
\begin{figure}
	\centering
	\includegraphics[width=\textwidth*2/3]{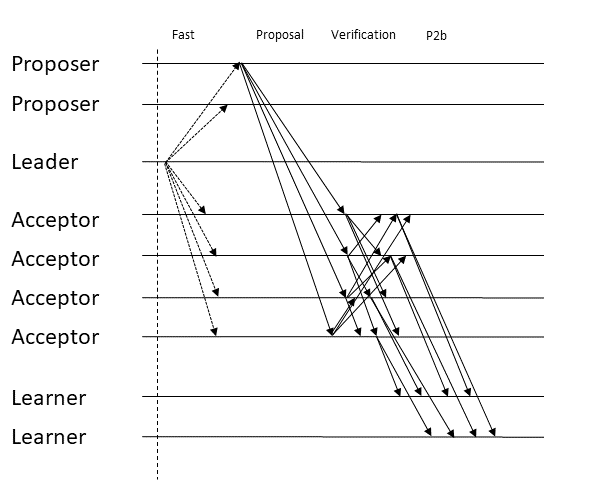}
	\caption{BGP's fast ballot message pattern}
	\label{bgp_fast}
\end{figure}
\begin{figure}
	\centering
	\includegraphics[width=\textwidth*2/3]{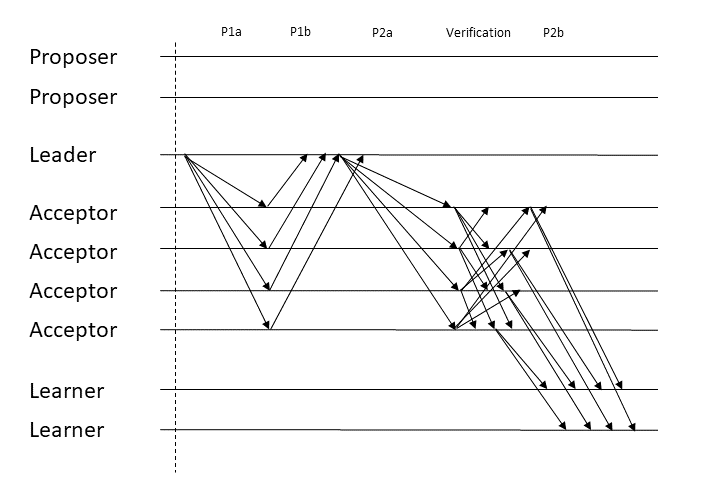}
	\caption{BGP's classic ballot message pattern}
	\label{bgp_classic}
\end{figure}

Next, we present the protocol for each type of ballot in detail. We start by describing fast ballots since their structure has consequences that implicate classic ballots. Figures \ref{bgp_fast} and \ref{bgp_classic} illustrate the message pattern for fast and classic ballots, respectively. In these illustrations, arrows that are composed of solid lines represent messages that can be sent multiple times per ballot (once per proposal) while arrows composed of dotted lines represent messages that are sent only once per ballot.

\begin{algorithm} 
	\caption{Byzantine Generalized Paxos - Acceptor a (view change)}
	\label{BFT-Proc}
	\textbf{Local variables:} $suspicions = \bot,\ new\_view = \bot,\ leader = \bot,\ view = 0, bal_a = 0,\ val_a = \bot,\ fast\_bal = \bot,\ checkpoint=\bot$
	\begin{algorithmic}[1]		
		\State \textbf{upon} \textit{suspect\_leader} \textbf{do} 
		\State\hspace{\algorithmicindent} \textbf{if} $suspicions[p] \neq true$ \textbf{then}
		\State\hspace{\algorithmicindent}\hspace{\algorithmicindent} $suspicions[p] = true$;
		\State\hspace{\algorithmicindent}\hspace{\algorithmicindent} $proof = \langle suspicion, view \rangle_{priv_a}$;
		\State \hspace{\algorithmicindent}\hspace{\algorithmicindent} $\Call{send}{SUSPICION, view,proof}$;	
		\State
		
		\State \textbf{upon} \textit{receive($SUSPICION, view_i, proof$)} from acceptor $i$ \textbf{do} 
		\State\hspace{\algorithmicindent} \textbf{if} $view_i \neq view$ \textbf{then}
		\State\hspace{\algorithmicindent}\hspace{\algorithmicindent} \textbf{return};
		\State\hspace{\algorithmicindent} \textbf{if} $proof_{pub_i} == \langle suspicion, view \rangle$ \textbf{then}
		\State\hspace{\algorithmicindent}\hspace{\algorithmicindent} $suspicions[i] = proof$;
		
		\State\hspace{\algorithmicindent} \textbf{if} $\#(suspicions) > f$ and $new\_view[view+1][p] == \bot$ \textbf{then}
		\State\hspace{\algorithmicindent}\hspace{\algorithmicindent} $change\_proof = \langle view\_change, view +1 \rangle_{priv_a}$;
		\State\hspace{\algorithmicindent}\hspace{\algorithmicindent} $new\_view[view+1][p] = change\_proof$;
		\State\hspace{\algorithmicindent}\hspace{\algorithmicindent} $\Call{send}{VIEW\_CHANGE, view+1, suspicions, change\_proof}$;
		\State
		
		\State\textbf{upon} \textit{receive($VIEW\_CHANGE, new\_view_i, suspicions, change\_proof_i$)} from acceptor $i$ \textbf{do} 
		\State\hspace{\algorithmicindent} \textbf{if} $new\_view_i \leq view$ \textbf{then}
		\State\hspace{\algorithmicindent}\hspace{\algorithmicindent}\textbf{return};
		\State
		\State\hspace{\algorithmicindent} $valid\_proofs = 0$;
		\State\hspace{\algorithmicindent} \textbf{for} $p$ \textbf{in} $acceptors$ \textbf{do} 
		\State\hspace{\algorithmicindent}\hspace{\algorithmicindent} $proof = suspicions[p]$;
		\State\hspace{\algorithmicindent}\hspace{\algorithmicindent} $last\_view = new\_view_i-1$;
		\State\hspace{\algorithmicindent}\hspace{\algorithmicindent} \textbf{if} $proof_{pub_p} == \langle suspicion, last\_view \rangle$ \textbf{then}
		\State\hspace{\algorithmicindent}\hspace{\algorithmicindent}\hspace{\algorithmicindent} $valid\_proofs \mathrel{+{=}} 1$;
		\State
		\State\hspace{\algorithmicindent} \textbf{if} $valid\_proofs \leq f$ \textbf{then}
		\State\hspace{\algorithmicindent}\hspace{\algorithmicindent} \textbf{return};
		\State
		\State\hspace{\algorithmicindent} $new\_view[new\_view_i][i] = change\_proof_i$;
		\State\hspace{\algorithmicindent} \textbf{if} $new\_view[view_i][a] == \bot$ \textbf{then}				
		\State\hspace{\algorithmicindent}\hspace{\algorithmicindent} $change\_proof = \langle view\_change, new\_view_i \rangle_{priv_a}$;
		\State\hspace{\algorithmicindent}\hspace{\algorithmicindent} $new\_view[view_i][a] = change\_proof$;
		\State\hspace{\algorithmicindent}\hspace{\algorithmicindent}  $\Call{send}{VIEW\_CHANGE, view_i, suspicions, change\_proof}$;
		\State
		\State\hspace{\algorithmicindent} \textbf{if} $\#(new\_view[new\_view_i]) \geq N-f$ \textbf{then}
		\State\hspace{\algorithmicindent}\hspace{\algorithmicindent} $view = view_i$;
		\State\hspace{\algorithmicindent}\hspace{\algorithmicindent} $leader = view\ mod\ N$;
		\State\hspace{\algorithmicindent}\hspace{\algorithmicindent} $suspicions = \bot$;
		\State\hspace{\algorithmicindent}\hspace{\algorithmicindent} $\Call{send}{LEADER, view, new\_view[view_i]}$ to leader;
	\end{algorithmic}
\end{algorithm}

\subsubsection{Fast Ballots} 

Fast ballots leverage the weaker specification of generalized consensus (compared to classic consensus) in terms of command ordering at different replicas, to allow for the faster execution of commands in some cases. The basic idea of fast ballots is that proposers contact the acceptors directly, bypassing the leader, and then the acceptors send their vote for the current sequence to the learners. If a conflict exists and progress isn't being made, the protocol reverts to using a classic ballot. This is where generalized consensus allows us to avoid falling back to this slow path, namely in the case where commands that ordered differently at different acceptors commute. \par 
However, this concurrency introduces safety problems even when a quorum is reached for some sequence. If we keep the original Fast Paxos message pattern~\cite{L06}, it's possible for one sequence $s$ to be learned at one learner $l_1$ while another non-commutative sequence $s'$ is learned before $s$ at another learner $l_2$. Suppose $s$ obtains a quorum of votes and is learned by $l_1$ but the same votes are delayed indefinitely before reaching $l_2$. In the next classic ballot, when the leader gathers a quorum of \textit{phase 1b} messages it must arbitrate an order for the commands that it received from the acceptors and it doesn't know the order in which they were learned. This is because, of the $N-f$ messages it received, $f$ may not have participated in the quorum and another $f$ may be Byzantine and lie about their vote, which only leaves one correct acceptor that participated in the quorum and a single vote isn't enough to determine if the sequence was learned or not. If the leader picks the wrong sequence, it would be proposing a sequence $s'$ that is non-commutative to a learned sequence $s$. Since the learning of $s$ was delayed before reaching $l_2$, $l_2$ could learn $s'$ and be in a conflicting state with respect to $l_1$, violating consistency. In order to prevent this, sequences accepted by a quorum of acceptors must be monotonic extensions of previous accepted sequences. Regardless of the order in which a learner learns a set of monotonically increasing sequences, the resulting state will be the same. The additional verification phase is what allows acceptors to prove to the leader that some sequence was accepted by a quorum. By gathering $N-f$ proofs for some sequence, an acceptor can prove that at least $f+1$ correct acceptors voted for that sequence. Since there are only another $2f$ acceptors in the system, no other non-commutative value may have been voted for by a quorum. \par
An interesting alternative to requiring $N-f$ proofs from each acceptor, would be for the leader to wait for $2f+1$ matching \textit{phase 1b} messages. Since at least $f+1$ of those would be correct, only that sequence could've been learned since any other non-commutative sequence would obtain at most $2f$ votes. Zyzzyva uses a similar approach of waiting for $3f+1$ to commit requests in single round-trip in executions where no faults occur~\cite{Kotla:2008}. However, this approach is unsuitable for BGP since it's possible for a sequence to be chosen by a quorum without the leader being aware of more than $f+1$ votes in its quorum. Since $f+1$ votes aren't enough to ensure the leader that the sequence was chosen by a quorum, the leader wouldn't be able to pick a learned sequence.\par
Next, we explain each of the protocol's steps for fast ballots in greater detail.

\noindent {\bf Step 1: Proposer to acceptors.}
To initiate a fast ballot, the leader informs both proposers and acceptors that the proposals may be sent directly to the acceptors. Unlike classic ballots, where the sequence proposed by the leader consists of the commands received from the proposers appended to previously proposed commands, in a fast ballot, proposals can be sent to the acceptors in the form of either a single command or a sequence to be appended to the command history. These proposals are sent directly from the proposers to the acceptors.\par

\begin{algorithm} 
	\caption{Byzantine Generalized Paxos - Acceptor a (agreement)}
	\label{BFT-Acc}
	\textbf{Local variables:} $leader = \bot,\ view = 0, bal_a = 0,\ val_a = \bot,\ fast\_bal = \bot,\ proven = \bot$
	\begin{algorithmic}[1]
		\State \textbf{upon} \textit{receive($P1A, ballot, view_l$)} from leader $l$ \textbf{do}
		\State \hspace{\algorithmicindent} \textbf{if} $view_l == view$ and $bal_a < ballot$ \textbf{then}
		\State \hspace{\algorithmicindent}\hspace{\algorithmicindent} $\Call{send}{P1B, ballot,bal_a,proven, val_a, proofs[bal_a]}$ to leader;
		\State \hspace{\algorithmicindent}\hspace{\algorithmicindent} $bal_a = ballot$;	
		\State \hspace{\algorithmicindent}\hspace{\algorithmicindent} $val_a = \bot$;

		\State
		\State \textbf{upon} \textit{receive($FAST,ballot,view_l$)} from leader \textbf{do}
		\State \hspace{\algorithmicindent} \textbf{if} $view_l == view$ \textbf{then}
		\State \hspace{\algorithmicindent}\hspace{\algorithmicindent} $fast\_bal[ballot] = true$;

		\State
		\State \textbf{upon} \textit{receive($VERIFY,view_i, ballot_i,val_i,proof$)} from acceptor $i$ \textbf{do}
		\State \hspace{\algorithmicindent} \textbf{if} $proof_{pub_i} == \langle ballot_i, val_i \rangle$ and $view == view_i$ \textbf{then}
		
		\State \hspace{\algorithmicindent}\hspace{\algorithmicindent} $proofs[ballot_i][val_i][i] = proof$;
		\State \hspace{\algorithmicindent}\hspace{\algorithmicindent} \textbf{if} $\#(proofs[ballot_i][val_i]) \geq N-f$ \textbf{then}
		\State \hspace{\algorithmicindent}\hspace{\algorithmicindent}\hspace{\algorithmicindent} $proven = val_i$;
		\State \hspace{\algorithmicindent}\hspace{\algorithmicindent}\hspace{\algorithmicindent} $\Call{send}{P2B, ballot_i, val_i, proofs[ballot_i][value_i]}$ to learners;
		
		\State
		\State \textbf{upon} \textit{receive$(P2A\_CLASSIC, ballot, view, value$)} from leader \textbf{do}
		\State \hspace{\algorithmicindent} \textbf{if} $view_l == view$ \textbf{then}
		\State \hspace{\algorithmicindent}\hspace{\algorithmicindent} $\Call{phase\_2b\_classic}{ballot, value}$; 
		
		\State		
		\State \textbf{upon} \textit{receive($P2A\_FAST, value$)} from proposer \textbf{do}
		\State \hspace{\algorithmicindent} $\Call{phase\_2b\_fast}{value}$;
	
		\State
		\Function{phase\_2b\_classic}{$ballot, value$}
		\State $univ\_commut = \Call{isUniversallyCommutative}{val_a}$;
		\State \textbf{if} $ballot \geq bal_a$ and $val_a ==  \bot$ and \ $!fast\_bal[bal_a]$ and ($univ\_commut$ or $proven == \bot$ or $proven == \Call{subsequence}{value, 0, \#(proven)}$) \textbf{then}
		\State \hspace{\algorithmicindent} $bal_a = ballot$;
		\State \hspace{\algorithmicindent} \textbf{if} $univ\_commut$ \textbf{then}
		\State \hspace{\algorithmicindent}\hspace{\algorithmicindent} $\Call{send}{P2B,bal_a, value}$ to learners;
		\State \hspace{\algorithmicindent} \textbf{else} 
		\State \hspace{\algorithmicindent}\hspace{\algorithmicindent} $val_a = value$;
		\State \hspace{\algorithmicindent}\hspace{\algorithmicindent} $proof = \langle ballot, val_a \rangle_{priv_a}$;
		\State \hspace{\algorithmicindent}\hspace{\algorithmicindent} $proofs[ballot][val_a][a] = proof$;
		\State \hspace{\algorithmicindent}\hspace{\algorithmicindent} $\Call{send}{VERIFY, view, ballot, val_a, proof}$ to acceptors;
		\EndFunction
		
		\State
		\Function{phase\_2b\_fast}{$ballot, value$}
		\State \textbf{if} $ballot == bal_a$ and $fast\_bal[bal_a]$ \textbf{then}
		\State \hspace{\algorithmicindent} \textbf{if} $\Call{isUniversallyCommutative}{value}$ \textbf{then}
		\State \hspace{\algorithmicindent}\hspace{\algorithmicindent} $\Call{send}{P2B,bal_a, value}$ to learners;
		\State \hspace{\algorithmicindent} \textbf{else}
		\State \hspace{\algorithmicindent}\hspace{\algorithmicindent} $val_a = val_a \bullet value$;
		\State \hspace{\algorithmicindent}\hspace{\algorithmicindent} $proof = \langle ballot, val_a \rangle_{priv_a}$;
		\State \hspace{\algorithmicindent}\hspace{\algorithmicindent} $proofs[ballot][val_a][a] = proof$;
		\State \hspace{\algorithmicindent}\hspace{\algorithmicindent} $\Call{send}{VERIFY, view, ballot, val_a, proof}$ to acceptors;
		\EndFunction
	\end{algorithmic}
\end{algorithm}

\noindent {\bf Step 2: Acceptors to acceptors.}
Acceptors append the proposals they receive to the proposals they have previously accepted in the current ballot and broadcast the resulting sequence and the current ballot to the other acceptors, along with a signed tuple of these two values. Intuitively, this broadcast corresponds to a verification phase where acceptors gather proofs that a sequence gathered enough support to be committed. This proofs will be sent to the leader in the subsequent classic ballot in order for it to pick a sequence that preserves consistency. To ensure safety, correct learners must learn non-commutative commands in a total order. When an acceptor gathers $N-f$ proofs for equivalent values, it proceeds to the next phase. That is, sequences do not necessarily have to be equal in order to be learned since commutative commands may be reordered. Recall that a sequence is equivalent to another if it can be transformed into the second one by reordering its elements without changing the order of any pair of non-commutative commands (in the pseudocode, proofs for equivalent sequences are being treated as belonging to the same index of the \emph{proofs} variable, to simplify the presentation). By requiring $N-f$ votes for a sequence of commands, we ensure that, given two sequences where non-commutative commands are differently ordered, only one sequence will receive enough votes even if $f$ Byzantine acceptors vote for both sequences. Outside the set of (up to) $f$ Byzantine acceptors, the remaining $2f+1$ correct acceptors will only vote for a single sequence, which means there are only enough correct processes to commit one of them. Note that the fact that proposals are sent as extensions to previous sequences is critical to the safety of the protocol. In particular, since the votes from acceptors can be reordered by the network before being delivered to the learners, if these values were single commands, it would be impossible to guarantee that non-commutative commands would be learned in a total order. \par
\noindent {\bf Step 3: Acceptors to learners.} Similarly to what happens in classic ballots, the fast ballot equivalent of the \textit{phase 2b} message, which is sent from acceptors to learners, contains the current ballot number, the command sequence and the $N-f$ proofs gathered in the verification round. One could think that, since acceptors are already gathering proofs that a value will eventually be committed, learners are not required to gather $N-f$ votes and they can wait for a single \textit{phase 2b} message and validate the $N-f$ proofs contained in it. However, this is not the case due to the possibility of learners learning sequences without the leader being aware of it. If we allowed the learners to learn after witnessing $N-f$ proofs for just one acceptor then that would raise the possibility of that acceptor not being present in the quorum of \textit{phase 1b} messages. Therefore, the leader wouldn't be aware that some value was proven and learned. The only way to guarantee that at least one correct acceptor will relay the latest proven sequence to the leader is by forcing the learner to require $N-f$ \textit{phase 2b} messages since only then will one correct acceptor be in the intersection of the two quorums. \par
\noindent {\bf Arbitrating an order after a conflict.} When, in a fast ballot, non-commutative commands are concurrently proposed, these commands may be incorporated into the sequences of various acceptors in different orders and, therefore, the sequences sent by the acceptors in \textit{phase 2b} messages will not be equivalent and will not be learned. In this case, the leader subsequently runs a classic ballot and gathers these unlearned sequences in \textit{phase 1b}. Then, the leader will arbitrate a single serialization for every previously proposed command, which it will then send to the acceptors. Therefore, if non-commutative commands are concurrently proposed in a fast ballot, they will be included in the subsequent classic ballot and the learners will learn them in a total order, thus preserving consistency.

\subsubsection{Classic Ballots} 

Classic ballots work in a way that is very close to the original Paxos protocol~\cite{Lam98}. Therefore, throughout our description, we will highlight the points where BGP departs from that original protocol, either due to the Byzantine fault model, or due to behaviors that are particular to our specification of the consensus problem.\par

In this part of the protocol, the leader continuously collects proposals by assembling all commands that are received from the proposers since the previous ballot in a sequence (this differs from classic Paxos, where it suffices to keep a single proposed value that the leader attempts to reach agreement on). When the next ballot is triggered, the leader starts the first phase by sending \textit{phase 1a} messages to all acceptors containing just the ballot number. Similarly to classic Paxos, acceptors reply with a \textit{phase 1b} message to the leader, which reports all sequences of commands they voted for. In classic Paxos, acceptors also promise not to participate in lower-numbered ballots, in order to prevent safety violations~\cite{Lam98}. However, in BGP this promise is already implicit, given (1) there is only one leader per view and it is the only process allowed to propose in a classic ballot and (2) acceptors replying to that message must be in the same view as that leader.

As previously mentioned, \textit{phase 1b} messages contain $N-f$ proofs for each learned sequence. By waiting for $N-f$ such messages, the leader is guaranteed that, for any learned sequence $s$, at least one of the messages will be from a correct acceptor that, due to the quorum intersection property, participated in the verification phase of $s$. Note that waiting for $N-f$ \textit{phase 1b} messages is not what makes the leader be sure that a certain sequence was learned in a previous ballot. The leader can be sure that some sequence was learned because each \textit{phase 1b} message contains cryptographic proofs from $2f+1$ acceptors stating that they would vote for that sequence. Since there are only $3f+1$ acceptors in the system, no other non-commutative sequence could've been learned. Even though each \textit{phase 1b} message relays enough proofs to ensure the leader that some sequence was learned, the leader still needs to wait for $N-f$ such messages to be sure that he is aware of any sequence that was previously learned. Note that, since each command is signed by the proposer (this signature and its check are not explicit in the pseudocode), a Byzantine acceptor can't relay made-up commands. However, it can omit commands from its \textit{phase 1b} message, which is why it's necessary for the leader to be sure that at least one correct acceptor in its quorum took part in the verification quorum of any learned sequence. \par
After gathering a quorum of $N-f$ \textit{phase 1b} messages, the leader initiates \textit{phase 2a} where it assembles a proposal and sends it to the acceptors. This proposal sequence must be carefully constructed in order to ensure all of the intended properties. In particular, the proposal cannot contain already learned non-commutative commands in different relative orders than the one in which they were learned, in order to preserve consistency, and it must contain unlearned proposals from both the current and the previous ballots, in order to preserve liveness (this differs from sending a single value with the highest ballot number as in the classic specification). Due to the importance and required detail of the leader's value picking rule, it will be described next in its own subsection. \par
The acceptors reply to \textit{phase 2a} messages by broadcasting their verification messages containing the current ballot, the proposed sequence and proof of their committal to that sequence. After receiving $N-f$ verification messages, an acceptor sends its \textit{phase 2b} messages to the learners, containing the ballot, the proposal from the leader and the $N-f$ proofs gathered in the verification phase. As is the case in the fast ballot, when a learner receives a \textit{phase 2b} vote, it validates the $N-f$ proofs contained in it. Waiting for a quorum of $N-f$ messages for a sequence ensures the learners that at least one of those messages was sent by a correct acceptor that will relay the sequence to the leader in the next classic ballot (the learning of sequences also differs from the original protocol in the quorum size, due to the fault model, and in that the learners would wait for a quorum of matching values instead of equivalent sequences, due to the consensus specification.)\par

\subsubsection{Leader Value Picking Rule} \textit{Phase 2a} is crucial for the correct functioning of the protocol because it requires the leader to pick a value that allows new commands to be learned, ensuring progress, while at the same time preserving a total order of non-commutative commands at different learners, ensuring consistency. The value picked by the leader is composed of three pieces: (1) the subsequence that has proven to be accepted by a majority of acceptors in the previous fast ballot, (2) the subsequence that has been proposed in the previous fast ballot but for which a quorum hasn't been gathered and (3) new proposals sent to the leader in the current classic ballot. \par
The first part of the sequence will be the largest of the $N-f$ proven sequences sent in the \textit{phase 1b} messages. The leader can pick such a value deterministically because, for any two proven sequences, they are either equivalent or one can be extended to the other. The leader is sure of this because for the quorums of any two proven sequences there is at least one correct acceptor that voted in both and votes from correct acceptors are always extensions of previous votes from the same ballot. If there are multiple sequences with the maximum size then they are equivalent (by same reasoning applied previously) and any can be picked. \par
The second part of the sequence is simply the concatenation of unproven sequences of commands in an arbitrary order. Since these commands are guaranteed to not have been learned at any learner, they can be appended to the leader's sequence in any order. Since $N-f$ \textit{phase 2b} messages are required for a learner to learn a sequence and the intersection between the leader's quorum and the quorum gathered by a learner for any sequence contains at least one correct acceptor, the leader can be sure that if a sequence of commands is unproven in all of the gathered \textit{phase 1b} messages, then that sequence wasn't learned and can be safely appended to the leader's sequence in any order. \par 
The third part consists simply of commands sent by proposers to the leader with the intent of being learned at the current ballot. These values can be appended in any order and without any restriction since they're being proposed for the first time.

\begin{algorithm}
	\caption{Byzantine Generalized Paxos - Learner l}
	\label{BFT-Learn}
	\textbf{Local variables:} $learned = \bot, messages = \bot$
	\begin{algorithmic}[1]			
		\State \textbf{upon} \textit{receive($P2B, ballot, value, proofs$)} from acceptor $a$ \textbf{do}
		\State \hspace{\algorithmicindent} $valid\_proofs = 0$;
		\State \hspace{\algorithmicindent} \textbf{for} $i$ \textbf{in} $acceptors$ \textbf{do}
		\State \hspace{\algorithmicindent}\hspace{\algorithmicindent} $proof = proofs[i]$;
		\State \hspace{\algorithmicindent}\hspace{\algorithmicindent} \textbf{if} $proof_{pub_i} == \langle ballot, value \rangle$ \textbf{then}
		\State \hspace{\algorithmicindent}\hspace{\algorithmicindent}\hspace{\algorithmicindent} 
		$valid\_proofs \mathrel{+{=}} 1$;
		\State
		\State \hspace{\algorithmicindent} \textbf{if} $valid\_proofs \geq N-f$ \textbf{then}
		\State \hspace{\algorithmicindent}\hspace{\algorithmicindent} $messages[ballot][value][a] = proofs$;
		\State
		\State \hspace{\algorithmicindent}\hspace{\algorithmicindent} \textbf{if} $\#(messages[ballot][value]) \geq N-f$ \textbf{then}
		\State \hspace{\algorithmicindent}\hspace{\algorithmicindent}\hspace{\algorithmicindent} $learned = \Call{merge\_sequences}{learned, value}$;

		\State
		\State \textbf{upon} \textit{receive($P2B, ballot, value$)} from acceptor $a$ \textbf{do}
		\State \hspace{\algorithmicindent} \textbf{if} $\Call{isUniversallyCommutative}{value}$ \textbf{then}
		\State \hspace{\algorithmicindent}\hspace{\algorithmicindent}
		 $messages[ballot][value][a] = true$;
		\State \hspace{\algorithmicindent}\hspace{\algorithmicindent} \textbf{if} $\#(messages[ballot][value]) > f$ \textbf{then} 
		\State \hspace{\algorithmicindent}\hspace{\algorithmicindent}\hspace{\algorithmicindent} $learned = learned \bullet value$;
		
		\State		
		\Function{merge\_sequences}{$old\_seq, new\_seq$}
		\State \textbf{for} $c$ \textbf{in} $new\_seq$ \textbf{do} 
		\State \hspace{\algorithmicindent} \textbf{if} $!\Call{contains}{old\_seq,c}$ \textbf{then}
		\State \hspace{\algorithmicindent}\hspace{\algorithmicindent}\hspace{\algorithmicindent} $old\_seq =  old\_seq \bullet c$;
		\State \textbf{return} $old\_seq$;
		\EndFunction
	\end{algorithmic}
\end{algorithm}

\subsubsection{Byzantine Leader}
The correctness of the protocol is heavily dependent on the guarantee that the sequence accepted by a quorum of acceptors is an extension of previous 
proven sequences. Otherwise, if the network rearranges \textit{phase 2b} messages such that they're seen by different learners in different orders, they will result in a state divergence. If, however, every vote is a prefix of all subsequent votes then, regardless of the order in which the sequences are learned, the final state will be the same. \par 
This state equivalence between learners is ensured by the correct execution of the protocol since every vote in a fast ballot is equal to the previous vote with a sequence appended at the end (Algorithm~\ref{BFT-Acc} lines \{43-46\}) and every vote in a classic ballot is equal to all the learned votes concatenated with unlearned votes and new proposals (Algorithm~\ref{BFT-Lead} lines \{42-45\}) which means that new votes will be extensions of previous proven sequences. However, this begs the question of how the protocol fares when Byzantine faults occur. In particular, the worst case scenario occurs when both $f$ acceptors and the leader are Byzantine (remember that a process can have multiple roles, such as leader and acceptor). In this scenario, the leader can purposely send \textit{phase 2a} messages for a sequence that is not prefixed by the previously accepted values. Coupled with an asynchronous network, this malicious message can be delivered before the correct votes of the previous ballot, resulting in different learners learning sequences that may not be extensible to equivalent sequences. \par
To prevent this scenario, the acceptors must ensure that the proposals they receive from the leader are prefixed by the values they have previously voted for. Since an acceptor votes for its $val_a$ sequence after receiving $N-f$ verification votes for an equivalent sequence and stores it in its $proven$ variable, the acceptor can verify that it is a prefix of the leader's proposed value (i.e., $proven \sqsubseteq value$). A practical implementation of this condition is simply to verify that the  subsequence of $value$ starting at the index $0$ up to index $length(proven)-1$ is equivalent to the acceptor's $proven$ sequence.

\subsection{Checkpointing}  BGP includes an additional feature that deals with the indefinite accumulation of state at the acceptors and learners. This is of great practical importance since it can be used to prevent the storage of commands sequences from depleting the system's resources. This feature is implemented by a special command $C^*$, proposed by the leader, which causes both acceptors and learners to safely discard previously stored commands. However, the reason why acceptors accumulate state continuously is because each new proven sequence must contain any previous proven sequence. This ensures that an asynchronous network can't reorder messages and cause learners to learn in different orders. In order to safely discard state, we must implement a mechanism that allows us to deal with reordered messages that don't contain the entire history of learned commands.\par
To this end, when a learner learns a sequence that contains a checkpointing command $C^*$ at the end, it discards every command in its $learned$ sequence except $C^*$ and sends a message to the acceptors notifying them that it executed the checkpoint for some command $C^*$. Acceptors stop participating in the protocol after sending \textit{phase 2b} messages with checkpointing commands and wait for $N-f$ notifications from learners. After gathering a quorum of notifications, the acceptors discard their state, except for the command $C^*$, and resume their participation in the protocol. Note that, since the acceptors also leave the checkpointing command in their sequence of proven commands, every valid subsequent sequence will begin with $C^*$. The purpose of this command is to allow a learner to detect when an incoming message was reordered. The learner can check the first position of an incoming sequence against the first position of its $learned$ and, if a mismatch is detected, it knows that either a pre and post-checkpoint message has been reordered. \par
When performing this check, two possible anomalies that can occur: either (1) the first position of the incoming sequence contains a $C^*$ command and the learner's $learned$ sequence doesn't, in which case the incoming sequence was sent post-checkpoint and the learner is missing a sequence containing the respective checkpoint command; or (2) the first position of the $learned$ sequence contains a checkpoint command and the incoming sequence doesn't, in which case the incoming sequence was assembled pre-checkpoint and the learner has already executed the checkpoint. \par
In the first case, the learner can simply store the post-checkpoint sequences until it receives the sequence containing the appropriate $C^*$ command at which point it can learn the stored sequences. Note that the order in which the post-checkpoint sequences are executed is irrelevant since they're extensions of each other. In the second case, the learner receives sequences sent before the checkpoint sequence that it has already executed. In this scenario, the learner can simply discard these sequences since it knows that it executed a subsequent sequence (i.e., the one containing the checkpoint command) and proven sequences are guaranteed to be extensions of previous proven sequences. \par
For brevity, this extension to the protocol isn't included in the pseudocode description.
\section{Correctness Proofs} \label{bft_proof}

This section argues for the correctness of the Byzantine Generalized Paxos protocol in terms of the specified consensus properties.\par

\begin{table}[h!]
	\renewcommand{\arraystretch}{1.5}
	\centering
	\begin{tabularx}{\linewidth}{ |c|X|}
		\hline
		Invariant/Symbol & Definition \\
		\hline
		$\thicksim$ & Equivalence relation between sequences \\
		\hline
		$X \overset{e}{\implies} Y$ & $X$ implies that $Y$ is eventually true \\
		\hline
		$X \sqsubseteq Y$ & The sequence $X$ is a prefix of sequence $Y$ \\
		\hline
		$\mathcal{L}$ & Set of learner processes \\
		\hline
		$\mathcal{P}$ & Set of proposals (commands or sequences of commands) \\
		\hline
		$\mathcal{B}$ & Set of ballots \\
		\hline
		$\bot$ & Empty command \\
		\hline		
		$learned_{l_i}$ & Learner $l_i$'s $learned$ sequence of commands \\
		\hline
		$learned(l_i,s)$ & $learned_{l_i}$ contains the sequence $s$ \\
		\hline
		$maj\_accepted(s,b)$ & $N-f$ acceptors sent phase 2b messages to the learners for sequence $s$ in ballot $b$ \\
		\hline
		$min\_accepted(s,b)$ & $f+1$ acceptors sent phase 2b messages to the learners for sequence $s$ in ballot $b$\\
		\hline
		$proposed(s)$ & A correct proposer proposed $s$ \\
		\hline
		
	\end{tabularx} 
	\vspace{\smallskipamount}
	\caption{BGP proof notation} 
	\label{table:bft_proof}
\end{table}

\subsubsection{Consistency}
\begin{theorem}
	At any time and for any two correct learners $l_i$ and $l_j$, $learned_{l_i}$ and $learned_{l_j}$ can subsequently be extended to equivalent sequences \par
\end{theorem} 
\textbf{Proof:} \par
\parbox{\linewidth-\algorithmicindent}{\strut1. At any given instant, $\forall s,s' \in \mathcal{P}, \forall l_i,l_j \in \mathcal{L}, learned(l_j,s) \land learned(l_i,s') \implies \exists \sigma_1,\sigma_2 \in \mathcal{P} \cup \{\bot\}, s \bullet \sigma_1 \thicksim s' \bullet \sigma_2$}  \par
\indent\indent\parbox{\linewidth}{\strut\textbf{Proof:} }\par
\indent\indent\indent\parbox{\linewidth-\algorithmicindent*3}{\strut1.1. At any given instant, $\forall s,s' \in \mathcal{P}, \forall l_i,l_j \in \mathcal{L}, learned(l_i,s) \land learned(l_j,s') \implies (maj\_accepted(s,b) \lor (min\_accepted(s,b) \land s \bullet \sigma_1 \thicksim x \bullet \sigma_2)) \land (maj\_accepted(s',b') \lor (min\_accepted(s',b') \land s' \bullet \sigma_1 \thicksim x \bullet \sigma_2)), \exists \sigma_1, \sigma_2 \in \mathcal{P} \cup \{\bot\}, \forall x \in \mathcal{P},\forall b,b' \in \mathcal{B}$} \par
\indent\indent\indent\indent\parbox{\linewidth-\algorithmicindent*4}{\strut\textbf{Proof:} A sequence can only be learned in some ballot $b$ if the learner gathers $N-f$ votes (i.e., $maj\_accepted(s,b)$), each containing $N-f$ valid proofs, or if it is universally commutative (i.e., $s \bullet \sigma_1 \thicksim x \bullet \sigma_2,\ \exists \sigma_1, \sigma_2 \in \mathcal{P} \cup \{\bot\}, \forall x \in \mathcal{P}$) and the learner gathers $f+1$ votes (i.e., $min\_accepted(s,b)$). The first case requires gathering $N-f$ votes from each acceptor and validating that each proof corresponds to the correct ballot and value (Algorithm \ref{BFT-Learn}, lines \{1-12\}). The second case requires that the sequence must be commutative with any other and at least $f+1$ matching values are gathered (Algorithm \ref{BFT-Learn}, \{14-18\}). This is encoded in the logical expression $s \bullet \sigma_1 \thicksim x \bullet \sigma_2$ which is true if the accepted sequence $s$ and any other sequence $x$ can be extended to an equivalent sequence, therefore making it impossible to result in a conflict.}

\indent\indent\indent\parbox{\linewidth-\algorithmicindent*3}{\strut1.2. At any given instant, $\forall s,s' \in \mathcal{P},\forall b,b' \in \mathcal{B}, maj\_accepted(s,b) \land maj\_accepted(s',b') \implies \exists \sigma_1,\sigma_2 \in \mathcal{P} \cup \{\bot\}, s \bullet \sigma_1 \thicksim s' \bullet \sigma_2$}\par
\indent\indent\indent\indent\parbox{\linewidth-\algorithmicindent*4}{\strut\textbf{Proof:} We divide the following proof in two main cases: (1.2.1.) sequences $s$ and $s'$ are accepted in the same ballot $b$ and (1.2.2.) sequences $s$ and $s'$ are accepted in different ballots $b$ and $b'$.}\par
\indent\indent\indent\indent\indent\parbox{\linewidth-\algorithmicindent*5}{\strut1.2.1.~At any given instant, $\forall s,s' \in \mathcal{P},\forall b \in \mathcal{B}, maj\_accepted(s,b) \land maj\_accepted(s',b) \implies \exists \sigma_1,\sigma_2 \in \mathcal{P} \cup \{\bot\}, s \bullet \sigma_1 \thicksim s' \bullet \sigma_2$} \par
\indent\indent\indent\indent\indent\indent\parbox{\linewidth}{\strut\textbf{Proof:} Proved by contradiction.}\par
\indent\indent\indent\indent\indent\indent\indent\parbox{\linewidth-\algorithmicindent*7}{\strut1.2.1.1.~At any given instant, $\forall s,s' \in \mathcal{P}, \forall \sigma_1,\sigma_2 \in \mathcal{P} \cup \{\bot \},\forall b \in \mathcal{B}, maj\_accepted(s,b) \land maj\_accepted(s',b) \wedge s \bullet \sigma_1 \not\thicksim s' \bullet \sigma_2$} \par
\indent\indent\indent\indent\indent\indent\indent\indent\parbox{\linewidth}{\strut\textbf{Proof:} Contradiction assumption.}\par
\indent\indent\indent\indent\indent\indent\indent\parbox{\linewidth-\algorithmicindent*7}{\strut1.2.1.2. Take a pair proposals $s$ and $s'$ that meet the conditions of 1.2.1 (and are certain to exist by the previous point), then $s$ and $s'$ contain non-commutative commands.}\par
\indent\indent\indent\indent\indent\indent\indent\indent\parbox{\linewidth-\algorithmicindent*8}{\strut\textbf{Proof:} The statement $\forall s,s' \in \mathcal{P}, \forall \sigma_1,\sigma_2 \in \mathcal{P} \cup \{\bot \}, s \bullet \sigma_1 \not\thicksim s' \bullet \sigma_2$ is trivially false because it implies that, for any combination of sequences and suffixes, the extended sequences would never be equivalent. Since there must be some $s,s',\sigma_1$ and $\sigma_2$ for which the extensions are equivalent (e.g., $s=s'$ and $\sigma_1=\sigma_2$), then the statement is false.}\par
\indent\indent\indent\indent\indent\indent\indent\parbox{\linewidth}{\strut1.2.1.3. A contradiction is found, Q.E.D. }\par
\indent\indent\indent\indent\indent\parbox{\linewidth-\algorithmicindent*5}{\strut1.2.2.~At any given instant, $\forall s,s' \in \mathcal{P},\forall b,b' \in \mathcal{B}, maj\_accepted(s,b) \land maj\_accepted(s',b') \land b \neq b' \implies \exists \sigma_1,\sigma_2 \in \mathcal{P} \cup \{\bot\}, s \bullet \sigma_1 \thicksim s' \bullet \sigma_2$} 
\indent\indent\indent\indent\indent\indent\parbox{\linewidth-\algorithmicindent*6}{\strut\textbf{Proof:} To prove that values accepted in different ballots are extensible to equivalent sequences, it suffices to prove that for any sequences $s$ and $s'$ accepted at ballots $b$ and $b'$, respectively, such that $b < b'$ then $s \sqsubseteq s'$. By Algorithm \ref{BFT-Acc} lines \{11-16,35,46\}, any correct acceptor only votes for a value in variable $val_a$ when it receives $2f+1$ proofs for a matching value. Therefore, we prove that a value $val_a$ that receives $2f+1$ verification messages is always an extension of a previous $val_a$ that received $2f+1$ verification messages. By Algorithm \ref{BFT-Acc} lines \{32,43\}, $val_a$ only changes when a leader sends a proposal in a classic ballot or when a proposer sends a sequence in a fast ballot.\strut}
\indent\indent\indent\indent\indent\indent\parbox{\linewidth-\algorithmicindent*6}{\strut In the first case, $val_a$ is substituted by the leader's proposal which means we must prove that this proposal is an extension of any $val_a$ that previously obtained $2f+1$ verification votes. By Algorithm \ref{BFT-Lead} lines \{24-39,41-47\}, the leader's proposal is prefixed by the largest of the proven sequences (i.e., $val_a$ sequences that received $2f+1$ votes in the verification phase) relayed by a quorum of acceptors in \textit{phase 1b} messages. Note that, the verification in Algorithm \ref{BFT-Acc} line \{27\} prevents a Byzantine leader from sending a sequence that isn't an extension of previous proved sequences. Since the verification phase prevents non-commutative sequences from being accepted by a quorum, every proven sequence in a ballot is extensible to equivalent sequences which means that the largest proven sequence is simply the most up-to-date sequence of the previous ballot. \strut}
\indent\indent\indent\indent\indent\indent\parbox{\linewidth-\algorithmicindent*6}{\strut To prove that the leader can only propose extensions to previous values by picking the largest proven sequence as its proposal's prefix, we need to assert that a proven sequence is an extension any previous sequence. However, since that is the same result that we are trying to prove, we must use induction to do so:\strut}
\indent\indent\indent\indent\indent\indent\indent\parbox{\linewidth-\algorithmicindent*7}{\strut\textbf{1. Base Case}: In the first ballot, any proven sequence will be an extension of the empty command $\bot$ and, therefore, an extension of the previous sequence.\strut}
\indent\indent\indent\indent\indent\indent\indent\parbox{\linewidth-\algorithmicindent*7}{\strut\textbf{2. Induction Hypothesis}: Assume that, for some ballot $b$, any sequence that gathers $2f+1$ verification votes from acceptors is an extension of previous proven sequences.\strut}
\indent\indent\indent\indent\indent\indent\indent\parbox{\linewidth-\algorithmicindent*7}{\strut\textbf{3. Inductive Step}: By the quorum intersection property, in a classic ballot $b+1$, the \textit{phase 1b} quorum will contain ballot $b$'s proven sequences. Given the largest proven sequence $s$ in the \textit{phase 1b} quorum (which, by our hypothesis, is an extension of any previous proven sequences), by picking $s$ as the prefix of its \textit{phase 2a} proposal (Algorithm \ref{BFT-Lead}, lines \{41-47\}), the leader will assemble a proposal that is an extension of any previous proven sequence.\strut}
\indent\indent\indent\indent\indent\indent\parbox{\linewidth-\algorithmicindent*6}{\strut In the second case, a proposer's proposal $c$ is appended to an acceptor's $val_a$ variable. By definition of the append operation, $val_a \sqsubseteq val_a \bullet c$ which means that the acceptor's new value $val_a \bullet c$ is an extension of previous ones.\par}
\indent\indent\indent\parbox{\linewidth-\algorithmicindent*3}{\strut1.3. For any pair of proposals $s$ and $s'$, at any given instant, $\forall x \in \mathcal{P}, \exists \sigma_1,\sigma_2,\sigma_3,\sigma_4 \in \mathcal{P} \cup \{\bot\}, \forall b,b' \in \mathcal{B}, (maj\_accepted(s,b) \lor (min\_accepted(s,b) \land s \bullet \sigma_1 \thicksim x \bullet \sigma_2)) \land (maj\_accepted(s',b') \lor (min\_accepted(s',b') \land s \bullet \sigma_1 \thicksim x \bullet \sigma_2)) \implies s \bullet \sigma_3 \thicksim s' \bullet \sigma_4$}\par
\indent\indent\indent\indent\parbox{\linewidth}{\strut\textbf{Proof:} By 1.2 and by definition of $s \bullet \sigma_1 \thicksim x \bullet \sigma_2$.}\par
\indent\indent\indent\parbox{\linewidth-\algorithmicindent*3}{\strut1.4. At any given instant, $\forall s,s' \in \mathcal{P}, \forall l_i,l_j \in \mathcal{L}, learned(l_i,s)\ \land\ learned(l_j,s') \implies \exists \sigma_1,\sigma_2 \in \mathcal{P} \cup \{\bot\}, s \bullet \sigma_1 \thicksim s' \bullet \sigma_2$ }\par
\indent\indent\indent\indent\parbox{\linewidth}{\strut\textbf{Proof:} By 1.1 and 1.3.}\par
\indent\indent\indent\parbox{\linewidth}{\strut1.5. Q.E.D. }\par
\parbox{\linewidth-\algorithmicindent*3}{\strut2. At any given instant, $\forall l_i,l_j \in \mathcal{L}, learned(l_j,learned_j) \land learned(l_i,learned_i) \implies \exists \sigma_1,\sigma_2 \in \mathcal{P} \cup \{\bot\}, learned_i \bullet \sigma_1 \thicksim learned_j \bullet \sigma_2$}\par
\indent\indent\parbox{\linewidth}{\strut\textbf{Proof:} By 1.}\par
\parbox{\linewidth}{\strut3. Q.E.D.} \par

\subsubsection{Nontriviality}
\begin{theorem}
If all proposers are correct, $learned_l$ can only contain proposed commands. \label{N-T1} \par
\end{theorem} 
\textbf{Proof:} \par
\parbox{\linewidth-\algorithmicindent}{\strut1. At any given instant, $\forall l_i \in \mathcal{L}, \forall s \in \mathcal{P}, learned(l_i,s) \implies \forall x \in \mathcal{P}, \exists \sigma \in \mathcal{P}, \forall b \in \mathcal{B}, \ maj\_accepted(s,b) \lor (min\_accepted(s,b) \land  (s \thicksim x \bullet \sigma \lor x \thicksim s \bullet \sigma))$ }\par
\indent\indent\parbox{\linewidth-\algorithmicindent*2}{\strut\textbf{Proof:} By Algorithm \ref{BFT-Acc} lines \{16,30,41\} and Algorithm \ref{BFT-Learn} lines \{1-18\}, if a correct learner learned a sequence $s$ at any given instant then either $N-f$ or $f+1$ (if $s$ is universally commutative) acceptors must have executed \textit{phase 2b} for $s$.}\par
\parbox{\linewidth}{\strut2. At any given instant, $\forall s \in \mathcal{P}, \forall b \in \mathcal{B}, maj\_accepted(s,b) \lor min\_accepted(s,b) \implies proposed(s)$ }\par
\indent\indent\parbox{\linewidth-\algorithmicindent*2}{\strut\textbf{Proof:} By Algorithm \ref{BFT-Acc} lines \{18-23\}, for either $N-f$ or $f+1$ acceptors to accept a proposal it must have been proposed by a proposer (note that the leader is considered a distinguished proposer).}\par
\parbox{\linewidth}{\strut3. At any given instant, $\forall s \in \mathcal{P}, \forall l_i \in \mathcal{L}, learned(l_i,s) \implies proposed(s)$}\par
\indent\indent\parbox{\linewidth}{\strut\textbf{Proof:} By 1 and 2.}\par
\parbox{\linewidth}{\strut4. Q.E.D.}\par

\subsubsection{Stability}
\begin{theorem}
If $learned_l = s$ then, at all later times, $s \sqsubseteq learned_l$, for any sequence $s$ and correct learner $l$\looseness=-1 \par
\end{theorem} 
\textbf{Proof:} By Algorithm \ref{BFT-Learn} lines \{12,18,20-26\}, a correct learner can only append new commands to its $learned$ command sequence.

\subsubsection{Liveness}
\begin{theorem}
For any proposal $s$ from a correct proposer, and correct learner $l$, eventually $learned_l$ contains $s$\par
\end{theorem} 
\parbox{\linewidth}{\textbf{Proof:}} \par
\parbox{\linewidth-\algorithmicindent}{\strut1. $\forall\ l_i \in \mathcal{L},\forall s,x \in \mathcal{P}, \exists \sigma \in \mathcal{P}, \forall b \in \mathcal{B}, maj\_accepted(s,b) \lor (min\_accepted(s,b) \land  (s \thicksim x \bullet \sigma \lor x \thicksim s \bullet \sigma))\overset{e}{\implies} learned(l_i,s)$}\par
\indent\indent\parbox{\linewidth-\algorithmicindent*2}{\strut\textbf{Proof:} By Algorithm \ref{BFT-Acc} lines \{10-15,28-29,41-42\} and Algorithm \ref{BFT-Learn} lines \{1-18\}, when either $N-f$ or $f+1$ (if $s$ is universally commutative) acceptors accept a sequence $s$ (or some equivalent sequence), eventually $s$ will be learned by any correct learner.}\par
\parbox{\linewidth-\algorithmicindent}{\strut2. $\forall s \in \mathcal{P}, proposed(s) \overset{e}{\implies} \forall x \in \mathcal{P}, \exists \sigma \in \mathcal{P}, \forall b \in \mathcal{B}, maj\_accepted(s,b) \lor (min\_accepted(s,b) \land  (s \thicksim x \bullet \sigma \lor x \thicksim s \bullet \sigma))$} \par
\indent\indent\parbox{\linewidth-\algorithmicindent*2}{\strut\textbf{Proof:} A proposed sequence is either conflict-free when its incorporated into every acceptor's current sequence or it creates conflicting sequences at different acceptors. In the first case, it's accepted by a quorum (Algorithm \ref{BFT-Acc}, lines \{10-15,28-29,41-42\}) and, in the second case, it's sent in \textit{phase 1b} messages to the in leader in the next ballot (Algorithm \ref{BFT-Acc}, lines \{1-4\}) and incorporated in the next proposal (Algorithm \ref{BFT-Lead}, lines \{24-47\}).} \par
\parbox{\linewidth}{\strut3. $\forall l_i \in \mathcal{L}, \forall s \in \mathcal{P}, proposed(s) \overset{e}{\implies} learned(l_i,s)$} \par
\indent\indent\parbox{\linewidth}{\strut\textbf{Proof:} By 1 and 2.} \par
\parbox{\linewidth}{\strut4. Q.E.D.}
\section{Conclusion and discussion}
\label{sec:disc}
We presented a simplified description of the Generalized Paxos specification and protocol, 
and an implementation of Generalized Paxos that is resilient against Byzantine faults.
We now draw some lessons and outline some extensions to our protocol that present interesting directions for future work and hopefully
a better understanding of its practical applicability.

\paragraph{Handling faults in the fast case.}
A result that was stated in the original Generalized Paxos
paper~\cite{Lamport2005} is that to tolerate $f$ crash faults and
allow for fast ballots whenever there are up to $e$ crash faults, the
total system size $N$ must uphold two conditions:
$N > 2f$ and $N > 2e+f$.
Additionally, the fast and classic quorums must be of size $N-e$ and $N-f$, respectively. This implies that there is a price to pay in terms of number of replicas and quorum size for being able to run fast operations during faulty periods.
An interesting observation from our work is that, since Byzantine fault tolerance already requires a total system size of $3f+1$ and a quorum size of $2f+1$, we are able to amortize the cost of both features, i.e., we are able to tolerate the maximum number of faults for fast execution without paying a price in terms of the replication factor and quorum size.

\paragraph{Extending the protocol to universally commutative commands.}
A downside of the use of commutative commands in the
context of Generalized Paxos is that the commutativity check is done
at runtime, to determine if non-commutative commands have
been proposed concurrently.
This raises the possibility of extending the protocol to handle
commands that are universally commutative, i.e., commute with every
other command. For these commands, it is known before executing them
that they will not generate any conflicts, and therefore it is not
necessary to check them against concurrently executing commands.  This
allows us to optimize the protocol by decreasing the number of phase
$2b$ messages required to learn to a smaller $f+1$ quorum. Since, by
definition, these sequences are guaranteed to never produce conflicts,
the $N-f$ quorum is not required to prevent learners from learning
conflicting sequences. Instead, a quorum of $f+1$ is sufficient to
ensure that a correct acceptor saw the command and will eventually
propagate it to a quorum of $N-f$ acceptors. 
This optimization is particularly useful in the context of 
geo-replicated systems, since it can be significantly faster
to wait for the $f+1$st message instead of the $N-f$th one.\par
The usefulness of this optimization is severely reduced if these sequences are processed like any other, by being appended to previous sequences at the leader and acceptors. New proposals are appended to previous proven sequences to maintain the invariant that subsequent proven sequences are extensions of previous ones. Since the previous proven sequences to which a proposal will be appended to are probably not universally commutative, the resulting sequence will not be as well. We can increase this optimization's applicability by sending these sequences immediately to the learners, without appending them to previously accepted ones. This special handling has the added benefit of bypassing the verification phase, resulting in reduced latency for the requests and less traffic generated per sequence. This extension can also be easily implemented by adding a single check in Algorithm \ref{BFT-Lead} lines \{19-20\}, Algorithm \ref{BFT-Acc} lines \{29-30,40-41\} and Algorithm \ref{BFT-Learn} lines \{14-18\}.

\paragraph{Generalized Paxos and weak consistency.}
The key distinguishing feature of the specification of Generalized
Paxos~\cite{Lamport2005} is allowing learners to learn concurrent
proposals in a different order, when the proposals commute. This idea
is closely related to the work on weaker consistency models like eventual or
causal consistency~\cite{Ahamad1995}, or consistency models that mix
strong and weak consistency levels like RedBlue~\cite{Li2012}, which attempt
to decrease the cost of executing operations by reducing coordination
requirements between replicas. 
The link between the two models becomes clearer with the introduction of 
universally commutative commands in the previous paragraph.
In the case of weakly consistent replication,
weakly consistent requests can be executed as if they were universally
commutative, even if in practice that may not be the case. E.g., checking 
the balance of a bank account and making a deposit do not commute since
the output of the former depends on their relative order. However,
some systems prefer to run both as weakly consistent operations, even
though it may cause executions that are not explained by a sequential
execution, since the semantics are still acceptable given
that the final state that is reached is the same and no invariants 
of the application are violated~\cite{Li2012}.




%

\vspace{1em}

\begin{normalsize}
\noindent {\bf Acknowledgements.} This work was supported by the European Research Council (ERC-2012-StG-307732) and FCT (UID/CEC/50021/2013).
\end{normalsize}

\bibliographystyle{plain}
\bibliography{references}


\end{document}